\documentclass[10pt]{article}

\usepackage{a4wide}
\usepackage{epsfig}
\usepackage{amsmath,amssymb}

\title{Unbiased computation of transition times by pathway recombination}

\author
{J.\ Kuipers,$^{1}$ G.T.\ Barkema$^{1,2}$\\
\\
\normalsize{$^{1}$Institute for Theoretical Physics, Utrecht University,}\\
\normalsize{Leuvenlaan 4, 3584 CE, Utrecht, The Netherlands}\\
\normalsize{$^{2}$Instituut-Lorentz for Theoretical Physics, Leiden University,}\\
\normalsize{Niels Bohrweg 2, 2333 CA, Leiden, The Netherlands}\\
}

\date{}

\begin{document}
\maketitle

\begin{abstract}
In many systems, the time scales of the microscopic dynamics and
macroscopic dynamics of interest are separated by many orders of
magnitude. Examples abound, for instance nucleation, protein folding,
and chemical reactions. For these systems, direct simulation of phase
space trajectories does not efficiently determine most physical
quantities of interest. The last decade has seen the advent of methods
circumventing brute force simulation. For most dynamical quantities,
these methods all share the drawback of systematical errors. We
present a novel method for generating ensembles of phase space
trajectories. By sampling small pieces of these trajectories in
different phase space domains and piecing them together in a smart way
using equilibrium properties, we obtain physical quantities such as
transition times. This method does not have any systematic error and
is very efficient; the computational effort to calculate the first
passage time across a free energy barrier does not increase with the
height of the barrier. The strength of the method is shown in the
Ising model.  Accurate measurements of nucleation times span almost
ten orders of magnitude and reveal corrections to classical nucleation
theory.

\end{abstract}

\section{Introduction}
The average time it takes a protein to fold, an undercooled liquid to
crystallize, or a chemically active molecule to react, can in
principle be obtained from brute force computer simulations, by simply
starting several times in the unfolded, liquid or pre-reaction state
and then integrating the dynamical equations in time until folding,
crystallization or reaction takes place. However, the typical time
scales of the microscopic dynamics and macroscopic dynamics of
interest are often separated by many orders of magnitude; for such
systems, direct simulation of relevant phase space trajectories is
very inefficient, if at all possible. In the last decade, methods have
been developed that sample transition pathways while circumventing
brute force simulation, such as transition path sampling~\cite{TPS},
transition interface sampling~\cite{TIS} and
milestoning~\cite{milestone}, but most dynamical quantities, e.g.\ the
average transition time, are not provided by those methods or are
provided with systematic errors. We present a method to determine such
dynamical quantities, free from systematic errors and very
efficient. Our method is generally applicable to systems with known
equilibrium properties, consisting of two regions with locally stable
states, separated from each other by a barrier, that may be very high.

\section{Method}
\subsection{Problem}
A typical question that can be addressed by our method is the
following: if a system is in equilibrium in a region A of the phase
space, what is the average time of first arrival in another region B?
One should typically think of A and B as attracting basins in phase
space, separated by a barrier. Examples of structures where such
phenomena are found include nucleation, protein folding and chemical
reactions. The simplest approach to answer such questions is by direct
simulation: the system is started in A and evolved in time until B is
reached, and this procedure is repeated many times to collect
statistics. If, however, after leaving A returns to it are much more
frequent than traversals to B, this direct simulation approach
becomes very inefficient, since most of the computational effort is
invested in dynamical trajectories from A back to A, rather than
to B. Our method distributes the computational effort more
efficiently, spending more time on actual traversals.

\subsection{Sampling subtrajectories}
\begin{figure}
\centering
\includegraphics[width=\textwidth]{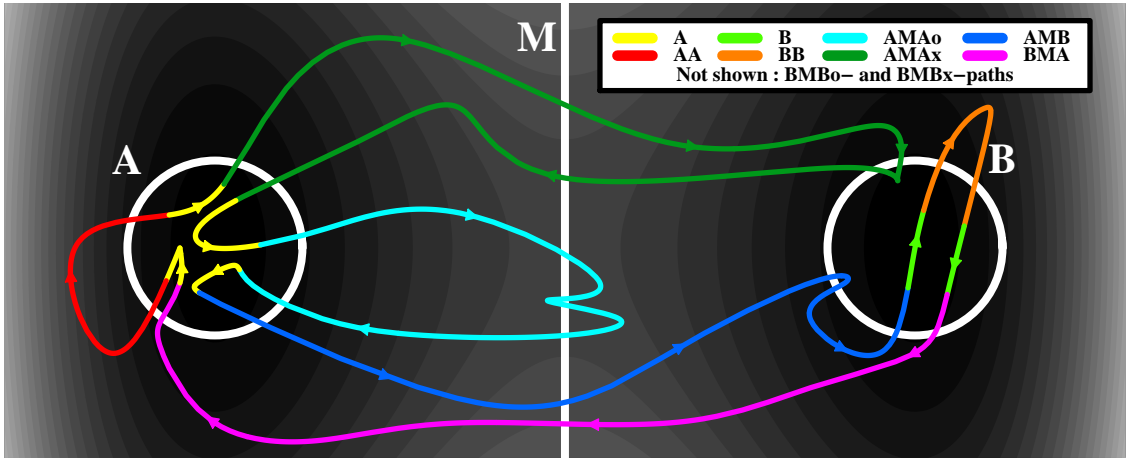}
\caption{Division of a trajectory in subtrajectories. A cut is made in
the trajectory every time that the system resides in region A or B longer
than some time $\tau_c$, which is the time in which the system loses
its memory of when, how and where it entered. Random recombinations of
these subtrajectories are equally valid trajectories.}
\label{fig::paths}
\end{figure}

The main idea behind our method is to sample different relatively
small pieces of the phase space trajectories, which we call {\it
subtrajectories}, and combine them with an appropriate weighting into
complete trajectories.  To classify the different subtrajectories, one
should identify a slice M in phase space, such that every path
connecting A and B has to pass through it. In the case of nucleation,
for instance, the regions A, B and M could be the set of states in
which the size of the largest nucleus is smaller than half, larger
than twice, or equal to the critical nucleus size (or a good estimate
of this). A long simulation trajectory can then be divided into
subtrajectories, which are classified according to (1) their initial
and final state (A or B) where the system resides at least the
correlation time $\tau_c$\footnote{The correlation time can be
different for the regions A and B, but for simplicity we take a single
value for $\tau_c$.}, (2) whether M has been crossed (denoted by an
`M' between the initial and final state), and (3) whether the path
crosses the other region without residing there longer than $\tau_c$
continuously or not (denoted by an `x' or `o' respectively).  All
these different types of subtrajectories are illustrated in figure
\ref{fig::paths}. The figure also shows A- and B-subtrajectories that
account for the time spent in A or B in excess of $\tau_c$.

The different types of subtrajectories are generated by performing
three different simulations. First, by starting in A, staying there
for a time $\tau_c$ and evolving from then, ignoring the paths that go
through M, the A- and AA-subtrajectories are sampled. Next, analogous
simulations around B are performed to find the B- and
BB-subtrajectories. Finally, simulations starting in M are performed
to generate the subtrajectories through M. These subtrajectories
through M are sampled by starting in M and evolving both forward and
backward in time until a correlation time $\tau_c$ is spent in A or B
continuously at both ends of the path. As shown in~\cite{sampleM1,sampleM2},
the starting points should be chosen from the set of
points of first arrival in M from either A or B to sample the
trajectories with the correct frequency. Because of time reversal
symmetry, this can be achieved by sampling points on M from its
equilibrium distribution, generating the trajectory by going forward
and backward in time, and ignoring all paths that encounter M again on
its backward part.

\subsection{Recombination of subtrajectories}

Now that the different types of subtrajectories are sampled, they
should be recombined to generate complete phase space trajectories.
These are concatinations of the subtrajectories, where the ones of the
types AA, AMAo, AMAx, AMB, BB, BMBo, BMBx and BMA are intertwined with
A- and B-subtrajectories. Recombination is based on the notion that in
the stable regions in phase space A and B the system will mostly
wander for a long time. If this time exceeds some value $\tau_c$ the
system has lost memory of where it entered. Since after this time
there is no correlation between the entrance and exit point of region
A or B, any random recombination of these subtrajectories would
constitute a valid trajectory. The subtrajectories of different
simulations can be recombined if given the proper weights; that is why
the method is efficient. In a straightforward simulation the
subtrajectories crossing M typically are extremely rare, which
strongly reduces statistical accuracy, but we generate additional
subtrajectories through M by starting there and these subtrajectories
can be combined with the AA- and BB-subtrajectories to sample long
trajectories.

The weights for the different sets of subtrajectories can be
determined from the condition that in the resulting long trajectories
the system must be found in A, B or M with the correct equilibrium
probabilities, $p^\text{(A)}$, $p^\text{(B)}$ and $p^\text{(M)}$. We
assume these are known, either analytically or from other kinds of
simulations, for instance parallel tempering~\cite{partemp}, methods
involving umbrella sampling~\cite{umbrella1,umbrella2}, cluster
algorithms~\cite{cluster1,cluster2} or the Wang-Landau
method~\cite{wanglandau}.  For each subtrajectory, we measure the time
it spends in regions A, B and M.  From these measurements we determine
for each class $C$ of subtrajectories (where $C$ can be AA, AMAo, AMAx,
AMB, BB, BMBo, BMBx and BMA) the average time spent in each region $X$
(where $X$ can be A, B or M) and these average times are also
determined for the class M of all subtrajectories crossing M; these
are the quantities $T_C^{(X)}$.  The notation $T_C$ without
superscript will denote the average total time spent on one
subtrajectory of class $C$.  Since these different subtrajectories are
always preceded and succeeded by either an A- or B-subtrajectory, the
time these latter subtrajectories take is also included in the times
$T_C^{(X)}$.  The frequency with which a long trajectory enters
subtrajectories of class $C$ is called $n_C$.  With these definitions,
it follows immediately that the probabilities of being in the three
regions satisfy the equalities

\begin{subequations}
\begin{align}
   p^\text{(A)} & =  n_\text{M} T_\text{M}^\text{(A)} + n_\text{AA} T_\text{AA}^\text{(A)},\\
   p^\text{(B)} & =  n_\text{M} T_\text{M}^\text{(B)} + n_\text{BB} T_\text{BB}^\text{(B)},\\
   p^\text{(M)} & =  n_\text{M} T_\text{M}^\text{(M)}.
\end{align}
\end{subequations}
From these the subtrajectory frequencies may be expressed as:
\begin{subequations}
\begin{align}
n_\text{AA}  & = (p^\text{(A)} - n_\text{M} T_\text{M}^\text{(A)}) / T_\text{AA}^\text{(A)},\\
n_\text{BB}  & = (p^\text{(B)} - n_\text{M} T_\text{M}^\text{(B)}) / T_\text{BB}^\text{(B)},\\
n_\text{M} & = p^\text{(M)} / T_\text{M}^\text{(M)}.
\end{align}
\label{eqn::nC}
\end{subequations}

During the simulations through M, the number of times $N_C$ that the
specific classes of subtrajectories through M are encountered, are
counted. These lead to the following relationships:
\begin{equation}
n_C = \frac{N_C}{N_\text{M}} \; n_\text{M},
\end{equation}
and these determine the remaining frequencies. 

From the frequencies of the different subtrajectories we immediately
obtain the probabilities for subtrajectories leaving A or B to be of
specific type:

\begin{subequations}
\begin{align}
p_\text{AA}   & = n_\text{AA}   / \mathcal{N}_A,\\
p_\text{AMAo} & = n_\text{AMAo} / \mathcal{N}_A,\\
p_\text{AMAx} & = n_\text{AMAx} / \mathcal{N}_A,\\
p_\text{AMB}  & = n_\text{AMB}  / \mathcal{N}_A,
\end{align}
\end{subequations}
with $\mathcal{N}_A={n_\text{AA}+n_\text{AMAo}+n_\text{AMAx}+n_\text{AMB}}$
and the analogous relations for the paths starting in B. Knowing these
probabilities we can randomly recombine subtrajectories with the proper
weights to generate complete trajectories.

\subsection{Determining transition times}
Depending on the exact quantity of interest, often the explicit
recombination process can be skipped and replaced by a combination of
appropriate averages over the subtrajectories. Here we specifically
want to address the average transition time from A to B. Note that a
traversal to B has to end with either an AMB- or
AMAx-subtrajectory. To calculate the average transition time, we need
to know the probabilities of finishing with an AMB- or
AMAx-subtrajectory, the average numbers of times the AA- and
AMAo-subtrajectories are traversed before this happens and the average
times these subtrajectories take. This results in the following
equation for the transition time:
\begin{equation}
T_\text{A$\rightarrow$B}^\ast =
  \frac{p_\text{AA} T_\text{AA} + p_\text{AMAo} T_\text{AMAo}
      + p_\text{AMAx} T_\text{AMAx}^{\text{first}}
      + p_\text{AMB} T_\text{AMB}^{\text{first}}}{p_\text{AMAx}+p_\text{AMB}}.
\end{equation}
The labels ``first'' are added to $T_\text{AMAx}$ and $T_\text{AMB}$,
since the first time that region B is reached is relevant for the
transition time, instead of the total time of the subtrajectory. These
can also be measured during the simulations.  We added an asterisk, to
distinguish these times from the first arrival time of B for a system
starting Boltzmann distributed in A; these times are the first arrival
times of B for a system starting in A with another distribution, namely
the distribution of points after being in A for a time $\tau_c$. To
obtain the real first arrival time $T_\text{A$\rightarrow$B}$, we
must perform final simulations that start Boltzmann distributed
in A until either $\tau_c$ time is spent in A, or arrival in B
occurs. We call the average time until a time $\tau_c$ in A is spent
$T_\text{A$\rightarrow$A}^\text{start}$, and the average time until
arrival in B occurs $T_\text{A$\rightarrow$B}^\text{start}$; these events
happen with the probabilities $p_\text{A$\rightarrow$A}^\text{start}$
and $p_\text{A$\rightarrow$B}^\text{start}$. The transition time from
A to B, starting Boltzmann distributed in A, is then
\begin{equation}
T_\text{A$\rightarrow$B} =
  p_\text{A$\rightarrow$A}^\text{start}
  \left(T_\text{A$\rightarrow$A}^\text{start} + T_\text{A$\rightarrow$B}^\ast\right)
  + p_\text{A$\rightarrow$B}^\text{start} T_\text{A$\rightarrow$B}^\text{start}.
\label{eqn::tAB}
\end{equation}

\subsection{Overview}
Since the determination of the transition time from A to B involves
multiple simulations in which a lot of quantities are measured, this
section provides an overview of the different simulations including
all measured quantities; these are presented in table
\ref{tab::overview}. With these quantities measured, equations
(\ref{eqn::nC}) - (\ref{eqn::tAB}) yield the average time of first
arrival in B for a system starting Boltmann distributed in A; other
dynamical quantities can be obtained as well from different related
equations.

\begin{table}[h!]
\centering
\begin{tabular}{c|l}
\multicolumn{2}{c}{Simulations starting with a time $\tau_c$ in A} \\
\hline
$T_\text{A}$             & The total time of an A-trajectory \\
$T_\text{AA}$            & The total time of an AA-trajectory \\
$T^\text{(A)}_\text{AA}$ & The time an AA-trajectory spends in A \\
\multicolumn{2}{c}{} \\
\multicolumn{2}{c}{Simulations starting with a time $\tau_c$ in B} \\
\hline
$T_\text{B}$             & The total time of a B-trajectory \\
$T_\text{BB}$            & The total time of a BB-trajectory \\
$T^\text{(B)}_\text{BB}$ & The time a BB-trajectory spends in B \\
\multicolumn{2}{c}{} \\
\multicolumn{2}{c}{Simulations starting in M} \\
\hline
$T_\text{M}$             & The total time of an M-trajectory \\
$T^\text{(A)}_\text{M}$, $T^\text{(B)}_\text{M}$, $T^\text{(M)}_\text{M}$ 
  & The time an M-trajectory spends in A, B or M, respectively \\
$T^\text{first}_\text{AMAx}$, $T^\text{first}_\text{AMB}$
  & The first time that an AMAx- or AMB-trajectory arrives in B \\
$N_\text{M}$ & The total number of M-trajectories \\
$N_\text{AMAx}$, $N_\text{AMAo}$, $N_\text{AMB}$
  & The number of AMAx-, AMAo- and AMB-trajectories, respectively \\
$N_\text{BMBx}$, $N_\text{BMBo}$, $N_\text{BMA}$
  & The number of BMBx-, BMBo- and BMA-trajectories, respectively \\
\multicolumn{2}{c}{} \\
\multicolumn{2}{c}{Simulations starting Boltzmann distributed in A} \\
\hline
$T^\text{start}_\text{A$\rightarrow$A}$
  & The time until a time $\tau_c$ is spent in A continuously \\
$T^\text{start}_\text{A$\rightarrow$B}$
  & The time to reach B without spending a time $\tau_c$ in A
continuously \\
$p^\text{start}_\text{A$\rightarrow$A}$, $p^\text{start}_\text{A$\rightarrow$B}$
  & The probabilities of starting with A$\rightarrow$A and A$\rightarrow$B
\end{tabular}
\caption{The measured quantities in the different simulations}
\label{tab::overview}
\end{table}

\section{Simple toy model}
\begin{figure}
\centering
\includegraphics[width=0.45\textwidth]{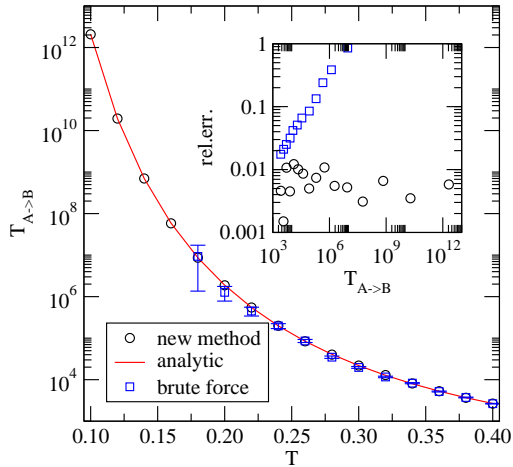}
\caption{Results of our method applied to the toy model. Shown are the
results of our method, compared to brute force simulations and an
analytic result. Error bars of our method are omitted since they are
very small. The inset shows the relative error of both our method and
brute force simulation.}
\label{fig::latticeresults}
\end{figure}

As a first test we applied this method to a system consisting of a
10$\times$10-lattice with a potential energy assigned to each
site\footnote{The potential energy of the lattice sites is given by
the function $V=((x+2y)(x+2y-2.6))^2 + (x-y-0.1)^2$ evaluated in the
points $x,y=0 \dots 0.9$ with steps of $0.1$. The sites $(0,0)$ and
$(0.9,0.9)$ form potential minima, while the diagonal in between is
around the potential maximum.}. The dynamics consist of jumps of a
single particle to neighboring sites with Metropolis~\cite{metropolis}
jump rates. ``Regions'' A and B are two opposing corners of the
lattice, which are minima of the potential. M is the diagonal in
between, which forms a ridge in the potential landscape.  The
simulations are performed for different values of the temperature.
Their results are shown in figure \ref{fig::latticeresults}, together
with the results of brute force simulations. Both simulations lasted
for one minute of CPU time. Also plotted is the exact transition time
that is calculated by solving a set of linear equations.

The results are as expected: both our method and the brute force
method sample the transition time without any systematic
error. However, with our method the statistical error is constant as a
function of the transition time, while the statistical errors of brute
force simulation increase proportional to the inverse square root of
the transition time, as the law of large numbers dictates (see the
inset of figure \ref{fig::latticeresults}).

\section{Ising model}
To show that the method does not only work efficiently in
low-dimensional toy systems, we apply it to determine the nucleation
time of a $64\times64$ Ising model with spin-flip dynamics and
Metropolis acceptance probabilities, for a large range of parameter
values for $\beta J$ and $\beta h$, i.e.\ the coupling constant and
the external magnetization in units of $k_\text{B}T$.

\begin{figure}[h]
\centering
\includegraphics[width=0.45\textwidth]{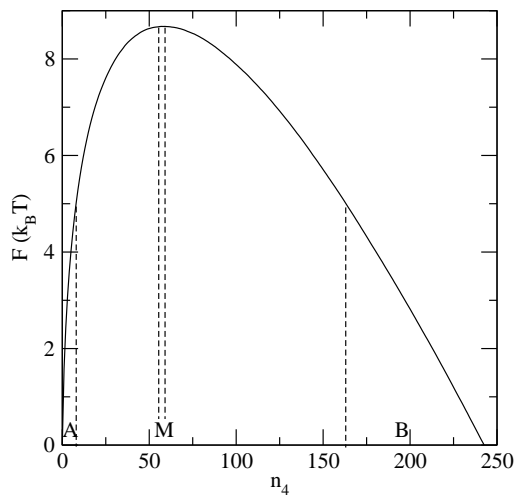}
\caption{Free energy of the two-dimensional Ising model with $64\times 64$ sites,
a coupling constant of $\beta J=0.60$ and an external field of $\beta h=0.06$,
as function of the number $n_4$ of spins parallel to the external field which
have four aligned neighbors. Also the pre-nucleation region A, the barrier region M
and the post-nucleation region B are indicated.}
\label{fig::freeenergy}
\end{figure}

Details of the simulations are as follows: we start in a metastable
state consisting of spins antiparallel to the external
magnetization. The coordinate used to characterize the regions A, B
and M is the number $n_4$ of spins that are parallel to the external
magnetization, and that have four neighbours which are also parallel
to it. This turns out to be a much better reaction coordinate than for
instance the size of the largest cluster of parallel spins. The
convenient property of $n_4$ is that it can only change by a maximum
of five; if the size of the largest cluster is taken as a reaction
coordinate, it can undergo large changes due to the merging or
splitting of clusters. To obtain the free energy as a function of
$n_4$ we use successive umbrella sampling~\cite{succumbrella}:
by restricting $n_4$ to either $i$ or $i+1$, the free energy
difference between $n_4=i$ and $n_4=i+1$ is determined; this process
is repeated for increasing values of $i$, until the free energy (as a
function of $n_4$) returns to the value at $n_4=0$. A typical result
of this free energy sampling is shown in figure \ref{fig::freeenergy}.

Next, the regions A, B and M are characterized in terms of $n_4$. The
region M is chosen at the top of the barrier and has width five, so
that every nucleation trajectory intersects it. The regions A and B
are such that the free energy barriers to its boundaries are
$5\;k_\text{B}T$. The motivation behind this is as follows: the barriers
should be high enough that the system will linger in A and B long
enough to lose correlation, but on the other hand low enough that it
can be crossed by thermal activation to sample AA- and
BB-trajectories; $5\;k_\text{B}T$ seems a reasonable choice for that. We will
call region A the pre-nucleation state and B the nucleated state,
since when the system arrives in B, it is extremely likely to continue
to a stable state in which most spins are aligned with the external
field. The regions A, B and M are also indicated in figure
\ref{fig::freeenergy}.  With the regions A, B and M defined and the
probabilities of being there known, the method can be applied to
determine the nucleation times. In our simulations, we took a
correlation time $\tau_c$ of 100 attemped spin flips per
site. Equilibration inside region A and B is fast, and the system is
certainly statistically uncorrelated within this time. We also
verified this in simulations with $\tau_c=200$ and 500.

\begin{figure}[h!]
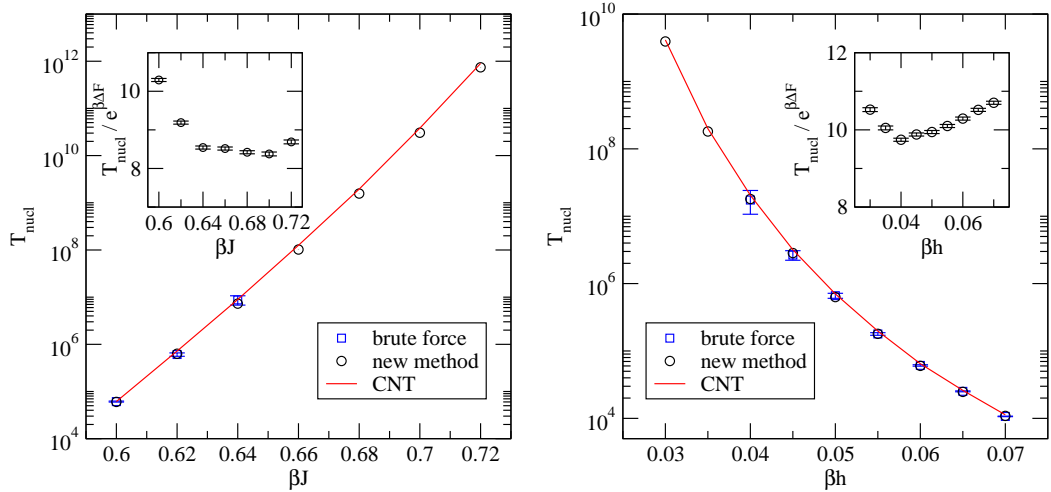

\vspace{1cm}
\begin{tabular}{cc}
\includegraphics[width=0.45\textwidth]{graph_bJ.eps}
&
\includegraphics[width=0.45\textwidth]{graph_bh.eps}
\end{tabular}
\caption{Nucleation times in the two-dimensional Ising model with
spin-flip dynamics, with a system of $64\times 64$ sites. Left: as
function of the coupling constant $\beta J$, for a fixed strength of
the external field $\beta h=0.06$. Right: as function of the strength
of the external field $\beta h$, for fixed coupling constant $\beta
J=0.6$. Error bars of our method are omitted since they are very
small. The insets show the deviations from classical nucleation
theory.}
\label{fig::ising}
\end{figure}

The resulting nucleation times are presented in figure
\ref{fig::ising}. Note that the nucleation times span 10 orders of
magnitude with constant relative statistical errors. For comparison,
results of brute force simulations (if possible) and classical nucleation
theory\footnote{Classical nucleation theory predicts that the nucleation
time equals $\text{const}\cdot\exp(\beta\Delta F_\text{max})$. The
constant is fitted to the data.} are also shown. The
general trend is well captured by CNT. However, as shown in the insets,
our method is accurate enough to reveal the shortcomings of CNT.

The computational effort in this calculation of the average nucleation
time is approximately 13 hours of CPU time on an AMD-64 single-processor
workstation for each set of temperature and field strength; one hour
is spent for the determination of the free energy as a function of
cluster size and 12 hours for the generation of the various
subtrajectories and the nucleation time. An equal amount of
computational effort was invested in the brute-force computations.

We would like to thank Henk van Beijeren for useful discussion.

\end{document}